\documentstyle[12pt]{article}
\begin{document}
\baselineskip=24pt

\begin{center}
{\large\bf A Device Structure for Electronic Transport Through Individual Molecules with Strong Coupling to Metallic Electrodes} 
\end{center}

\vspace{0.3in}

\begin{center}
{\bf Subhasis Ghosh} \\
School of Physical Sciences, Jawaharlal Nehru University,
New Delhi 110067
\end{center}

\begin{center}
{\bf Henny Halimun, Jaewon Choi, Saurabh Lodha and David Janes} \\

School of Electrical and Computer Engineering, Purdue University, West Lafayatte, Indiana 47907
\end{center}

\vspace{0.5in}

\begin{center}
{\bf Abstract}
\end{center}

We present a simple and reliable method for making electrical contacts to small organic molecules with thiol endgroups. 
Nanometer-scale gaps between metallic electrodes have been fabricated by passing a large current through a lithographically-patterned Au-line with appropriate thickness. 
Under appropriate conditions, the passage of current breaks the Au-line, creating two opposite facing electrodes separated by a gap comparable to the length of small organic molecules. Current-voltage characteristics have been measured both before and after deposition of short organic molecules. 
The resistance of single 1,4-benzenedithiol and 1,4-bezenedimethanedithiol molecules were found to be 9M$\Omega$ and 26M$\Omega$, respectively. 
The experimental results indicate strong electronic copuling to the contacts and are discussed using a relatively simple model of mesoscopic transport. 
The use of electrodes formed on an insulating surface by lithography and electromigration provides a stable structure suitable for integrated circuit applications.

\newpage

Recently, it has been demonstrated\cite{jc,cpc,dig,mar1} that a single or a small assembly of molecules with length scale$\sim$1nm can perform functions which are analogous to those of conventional semiconductor device elements. Molecular components are envisioned by many as a way to realize ultradense functional electronic circuits for logic and memory applications. One feature that distinguishes molecular electronics from the conventional electronic devices is the bottom-up method based on thermodynamically driven self-assembly of molecules, which may lead to molecule-based robust, fault-tolerant and defect-free nanodevices. In this context, metal-molecule-metal(m-M-m) junctions are being investigated for two reasons: first their application as key elements in future nanoelectronics and second the fundamental interest of understanding the mesoscopic transport through a single or small assembly of molecules. In spite of considerable progress, the field of molecule-based experiments is in its infancy. The major challenge in this field is interfacing nanometer-scale objects (molecules) with macroscopic circuits. The size of typical organic molecules is in the range of 1-10nm, hence at least two macroscopic metallic electrodes separated by the same distance are required for the most basic, nevertheless the most fundamental electrical measurements. Controlled fabrication in this length scale is beyond the scope of most present day lithographic techniques. 

Several methods have been employed to fabricate metallic electrodes whose separation is comparable to the length of the organic molecules. In the first approach\cite{mar2,ck,jr}, a mechanically controlled break junction(MCBJ) has been used to define electrodes with nanoscale separations suitable for transport measurements through a single molecule. In the second approach\cite{sd,djw,xdc}, a scanning probe microscope(SPM) has been used to study the mesoscopic transport through molecules, generally within an organic monolayer. A third approach\cite{hp,jp} used a novel method combining electron-beam lithography and electromigration to fabricate metallic electrodes with nanometer separation. All of these techniques have been used to interface a chemically synthesized nanostructure to macroscopic electrodes separated by $\sim$1nm to investigate the mesoscopic transport either in a single molecule or a small assembly of molecules. The first two methods generally lack the electrical and mechanical stability required for transport measurements under different ambient conditions, e.g. over wide temperature ranges. Subsequent analysis\cite{ege} has raised questions about whether individual molecules are actually bonded across the gap in break junctions fabricated by the methods proposed in Ref.\cite{mar2,ck,jr}. In principle, our method of fabrication of two electrodes with nanometer separation is similar to third method\cite{hp,jp}. Our method is relatively simple and highly reproducible and does not require sophisticated lithography. It is also suitable for performing transport measurements under different ambient conditions. 

In this letter, we present a simple method to fabricate metallic electrodes with nanometer separation, and subsequent binding of simple organic molecules in the gaps using appropriate molecular end groups. 
The fabrication has been achieved by passing a large electrical current through a thin Au-finger fabricated by photolithography. 
High current induces electromigration in the Au-finger and yields two stable electrodes with nanometer separations, which are ideally suitable for transport measurements through nanometer size molecules. 
Molecules of 1,4-benzedithiol(BDT) and 1,4-benzenedimethanedithiol(BDMT) have been chemisorbed by self-assembly technique between two electrodes in order to form m-M-m junctions. 
The use of the lithographic approach to form the structures allows the formation of suitable contact pads for probing or eventual integration into circuits.

A pattern consisting of a Au line of length 2-10$\mu$m and width 2$\mu$m with two contact pads was deposited on an oxidized Si substrate (SiO$_2$/Si). The thickness of the Au-finger was around 100$\AA$ and a thin wetting layer of Ti was used below the Au layer. These fabrication parameters are chosen such that the resistance of the Au-finger is within the range of 50 to 250$\Omega$, which has been found to be the optimum resistance value for fabrication of electrodes with nanometer separation. Fig.1 shows a representative I-V characteristic and corresponding conductance trace of a Au line during the breaking process. In this device, the conductance of the Au-finger remains constant at about 5mS until the onset of breaking at 10.5V, at which point the conductance drops abruptly to 1.5$\times$10$^{-5}$mS. The conductance following this procedure is due to the tunneling barrier between the nanometer separated electrodes. By calibrating the value of conductance of the break junction with that of metal-air-metal junction in MCBJ(Fig.5 in Ref.\cite{ck}), we have estimated the distance between the two electrodes to be around 1.2nm. 
The distance between two nanoelectrodes has been varied from 3.5nm to 0.8nm by this method, as inferred from the measured conductance after formation of the break. 
Following formation and initial electrical characterization of the break junctions, 1mM solutions of BDT or BDMT in dichloromethane or ethanol were used to deposit the desired molecules, via immersion of the whole chip or deposition of a droplet onto the break junction sample.
Ideally, the molecules with thiol end groups dock between two electrodes and a stable chemical bond between the sulphur atom and Au surface established, when the separation of the two electrodes matches with the length of the molecule. 
In order to verify the docking of molecular layers to Au surfaces, self assembled monolayer(SAM) were prepared on clean Au surfaces using the same immersion procedures, and were subsequently characterized by reflection absorption infrared(RAIR) spectroscopy. The spectra from the companion samples are shown in the insets of Figs. 2 and 3. For the BDT sample, the peaks at around 1500cm$^{-1}$ are due to benzene ring stretch(C=C-C) in BDT. The peaks at around 1000cm$^{-1}$ and 800cm$^{-1}$ are due to C-H in plane and out of plane bend in BDT, respectively. The peaks in the BDMT spectrum have similar origin as those in BDT.

Fig.2 and Fig.3 show the measured current-voltage(I-V) for BDT and BDMT, respectively. The figures also show the measured I-V of the respective contact structures before deposition of the molecules. I-V measurements were performed on more than 100 break junctions for each molecule. Generally, the conductance was observed to increase with increasing deposition time of the molecules; based on the width of the contacts, this is believed to be correspond to an increase in the number of molecules bridging the contact gap. The increased conductance was generally observed in contact gaps with gap widths (inferred from pre-molecule conductance) that were comparable to the molecular length, and not observed in contacts with significantly larger gaps. It should also be noted that the conductance did not systematically increase upon exposure to the organic solvent (without molecules), indicating that the observed conductance was not due to leakage paths along the oxide surface or solvent-induced reconfiguration of the contact structures. The data are highly reproducible and the break junctions are stable up to $\sim$10V, which appears to be significantly higher than the stable bias for MCBJ samples ($\sim$1.2V)\cite{jr}. 

We have observed that devices showing a significant increase in conductivity following deposition of molecules generally exhibit I-V characteristics with comparable shapes, but approximately six distinct current levels. 
Fig.4 shows that the I-V curves of 6 devices, each divided by an integer, along with a histogram indicating the number of devices which exhibited the various conductance values. The observance of curves with the same shape, but scaled by integers, indicates that the curves observed for N=1 likely correspond to devices in which a single BDMT molecule is responsible for the conduction.
Similar characteristics were obtained by Cui et. al.\cite{xdc} in case of octanedithiol molecule using conducting atomic force microscope. 
Note that the data shown in Fig. 2 corresponds to a case in which a number of molecules are thought to be conducting in parallel, while 
the data in Fig. 3 correspond to the lowest observed conductance in this series, and therefore is thought to correspond to a single molecule.
The I-V characteristics are highly reproducible, remain constant over several repeated measurements and are not affected by the factors mentioned in Ref.\cite{jr}

There are several notable features of the I-V characteristics of the samples containing BDT and BDMT molecules. 
The I-V characteristics are linear at low bias($\leq 0.3V$) and the measured zero bias resistance of single BDT and BDMT molecules are found to be 10M$\Omega$ and 26M$\Omega$, respectively.
As expected due to the presence of the -CH$_2$- groups, the observed resistance of the BDMT is higher. 
The significant conductance at low-bias, and lack of a significant conductance gap, 
are in strong contrast to prior reports of measurements obtained by SPM\cite{sd,wt} and by prior MCBJ studies. 
The measured values of low-field conductance are significantly higher than those observed in measurements using other contact structures. 
Over the measured voltage range, the I-V characteristics are relatively symmetric and nonlinear. 
This behavior, and the lack of a significant conductance gap, 
are consistent with the behavior expected for molecular conductors in which the
density of states in the molecule is broadened significantly by strong coupling
to the contacts\cite{wt, psd1, psd2}. The I-V characteristics of representative devices were measured versus temperature, and were observed to be consistent with tunneling-based conduction.

It should be noted that the I-V characteristics may be dominated by the contact properties, rather than intrinsic molecular properties, unless there are strongly coupled contacts at both ends of the molecule\cite{psd1,ege,psd2}, for example in chemically bonded contacts to each of the two opposite facing electrodes in a break junction,. 
In our case, the highly reproducible and symmetric I-V characteristics indicate that the contacts are stable and that the measured I-V relationships correspond to intrinsic electronic properties of individual molecules.
The superposition of all I-V traces on a fundamental curve(Fig.4) provides evidence of (a) stable Au-S contacts at both ends of the molecule and (b) negligible in plane molecule-molecule interactions. 
Although techniques are currently not available to image the molecules within the break junction,
the assumption of chemical bonding to the contacts is reasonable, based on the solvent-based deposition procedures on
the pre-formed structures. 
In previous studies\cite{mar2,ck,jr} on current transport through a single molecule involving MCBJ structures, 
contacts to the molecule are often made and broken by mechanically pulling apart the junction and
then bringing the contacts back together between series of measurements. 
This procedure likely results in variations in the microscopic contact at one end of the 
molecule due to either i) breaking and formation of a Au-S bond
or ii) formation of a physisorbed contact. 
These variations in the microscopic nature of the contact make it difficult to
achieve symmetric and reproducible I-V characteristics in MCBJ samples. 
In has been suggested \cite{ege} that the current observed in early MCBJ structures
flowed through an overlapping pair of molecules bonded between two opposite facing electrodes, rather then
through an individual molecule bonded to both contacts. 
In our experiment, it is likely that the Au-S bonds are stable at both ends of the molecule,
and therefore that the contacts remain stable versus time. 
In addition to providing consistent coupling to the contacts, 
this also provides consistent amounts of fractional charge to be transferred onto the molecule,
leading to consistent shifts of the molecular level and consistent charging energy during I-V measurements. 
Hence the origin of symmetric and reproducible I-V characteristics in our case may be different from those obtained using MCBJ\cite{mar2,ck,jr}.

Electrical transport through m-M-m junctions depends critically on the position of the Fermi level E$_F$, 
the broadening($\Gamma$) of the molecular levels $\epsilon_0$ due to the coupling to the metal contacts and the 
effect of a relatively large single electron charging energy U on the molecular levels. 
The relative strength of $\Gamma$ with respect to U dictates the nature of transport process through the m-M-m junction. 
If $U>\Gamma$, it is expected that transport will be dominated by a Coulomb blockade mechanism characterized by integral charge transfer. 
When molecules are chemisorbed between two electrodes with strong metal-molecule coupling ($U\leq\Gamma$), the transport is expected to be in a self consistent regime characterized by fractional charge transfer.
The absence of Coulomb blockade and observation of symmetric I-V characteristics, as shown in Fig.2 and Fig.3, indicate that the transport in this study is in the fractional charge transfer regime. 
In this case the conductance gap $E_{gap}\sim|E_F-\epsilon_0|$ will be reduced by the contact and thermal broadening and can be given by\cite{psd1} $E_{gap}\sim[(|E_F-\epsilon_0|)-(2\Gamma+4k_BT)]$.
Hence, strong broadening of molecular levels, associated with strong coupling to the contacts, and a large charging energy can account for the observed reduction of the conductance gap. 
These changes in the gap can explain, at least partially, the striking differences among the I-V characteristics measured by different techniques\cite{mar1,jr,sd,wt}. 
It has been observed that different measurements\cite{mar1,jr} using the same nominal technique(MCBJ) exhibit large variations in I-V characteristics of a single molecule due to subtle differences\cite{ege} in the contact formation between Au and S atoms.

In conclusion, we have presented a novel method to form electrodes with nanometer scale spacing and to dock single (or few) molecules between the electrodes. 
We have measured the current-voltage relationships for two short aromatic molecules and observed repeatable characteristics.
The lack of a significant conductance gap and the relatively large conductance values at low bias indicate
that strong coupling has been achieved between the molecule and the metallic contacts. 
This method allows electrical transport measurements through a single molecule and can be used to characterize a large number of important organic molecules.
The use of lithographically defined structures allows this approach to be used for integrated circuit applications.

The authors would like to thank Supriyo Datta for stimulating discussions. This work was supported in part by Army Research Office and by NASA.

\newpage

\noindent {\bf Figure Captions}

\vspace{0.3in}

\begin{description}

\item[Fig.1.] (a) A representative I-V characteristics of Au-finger during breaking process due to the passage of high current. (b) Corresponding conductance trace of the Au-finger during breaking process. Inset scheme of the experimental set-up, an organic molecule with two thiol endgroups(1,4-benzenedithiol) chemisorbed between two Au electrodes.

\vspace{0.2in}

\item[Fig.2.] Measured I-V characteristics of 1,4-benzene dithiol(BDT) at room temperature (open circles) and of contact structure before deposition of molecules (solid circles). Current through the break junction increases by almost three order of magnitude after docking the BDT molecule. Inset shows the RAIR of BDT on Au.

\vspace{0.2in}

\item[Fig.3.] Measured I-V characteristics of 1,4-benzenedimethanedithiol (BDMT) at room temperature, along with that of contact structure before deposition of molecules. Inset shows the RAIR of BDMT on Au. 
\vspace{0.2in}

\item[Fig.4.] Six representative I-V curves of BDMT from different groups, which are integer multiple of a fundamental curve. Curves are divided by 1,2,3,4,5 and 6. Inset shows the histogram of no. of occurrences with current divisor.

\end{description}


\newpage

\begin{thebibliography}{99}

\bibitem{jc} J. Chen, M. A. Reed, A. M. Rowlett and J. M. Tour, Science, {\bf 286}, 1550 (1999).

\bibitem{cpc} C. P. Collier et. al. Science, {\bf 285}, 391 (1999)

\bibitem{dig} D. I. Gittins, D. Bethell, D. J. Schiffrin and R. J. Nichols, Nature, {\bf 408}, 67 (2000)

\bibitem{mar1} M. A. Reed, J. Chen, A. M. Rawlett, D. W. Price, and J. M. Tour, Appl. Phys. Lett. {\bf 78}, 3735 (2001). 

\bibitem{mar2} M. A. Reed, C. Zhou, C. J. Muller, T. P. Burgin and J. M. Tour, Science, {\bf 278}, 252 (1997).

\bibitem{ck} C. K. Kergueris et. al. Phys. Rev. B. {\bf 59}, 12505 (1999).

\bibitem{jr} J. Reichert et. al. Phys. Rev. Lett. {\bf 88}, 176804 (2002).

\bibitem{sd} S. Datta et. al., Phys. Rev. Lett. {\bf 79}, 2530 (1997).

\bibitem{djw} D. J. Wold and C. D. Frisbie, J. Am. Chem. Soc. {\bf 123}, 5549 (2001).

\bibitem{xdc} X. D. Cui et. al. Science, {\bf 294}, 571 (2001).

\bibitem{hp} H. P. Park et. al. Appl. Phys. Lett. {\bf 75}, 301 (1999).

\bibitem{jp} J. Park et. al. Nature, {\bf 417}, 722 (2002).

\bibitem{ege} E. G. Emberley and G. Kirczenow, Phys. Rev. B {\bf 64}, 235412 (2001).


\bibitem{wt} W. Tian et. al. J. Chem. Phys. {\bf 109}, 2874 (1998)

\bibitem{psd1} P. S. Damle, A. W. Ghosh, and S. Datta, Chem. Phys. {\bf 281}, 171 (2002).


\bibitem{psd2} P. S. Damle, A. W. Ghosh, and S. Datta, Phys. Rev. B {\bf 64}, 201403 (2001).


\end{thebibliography}
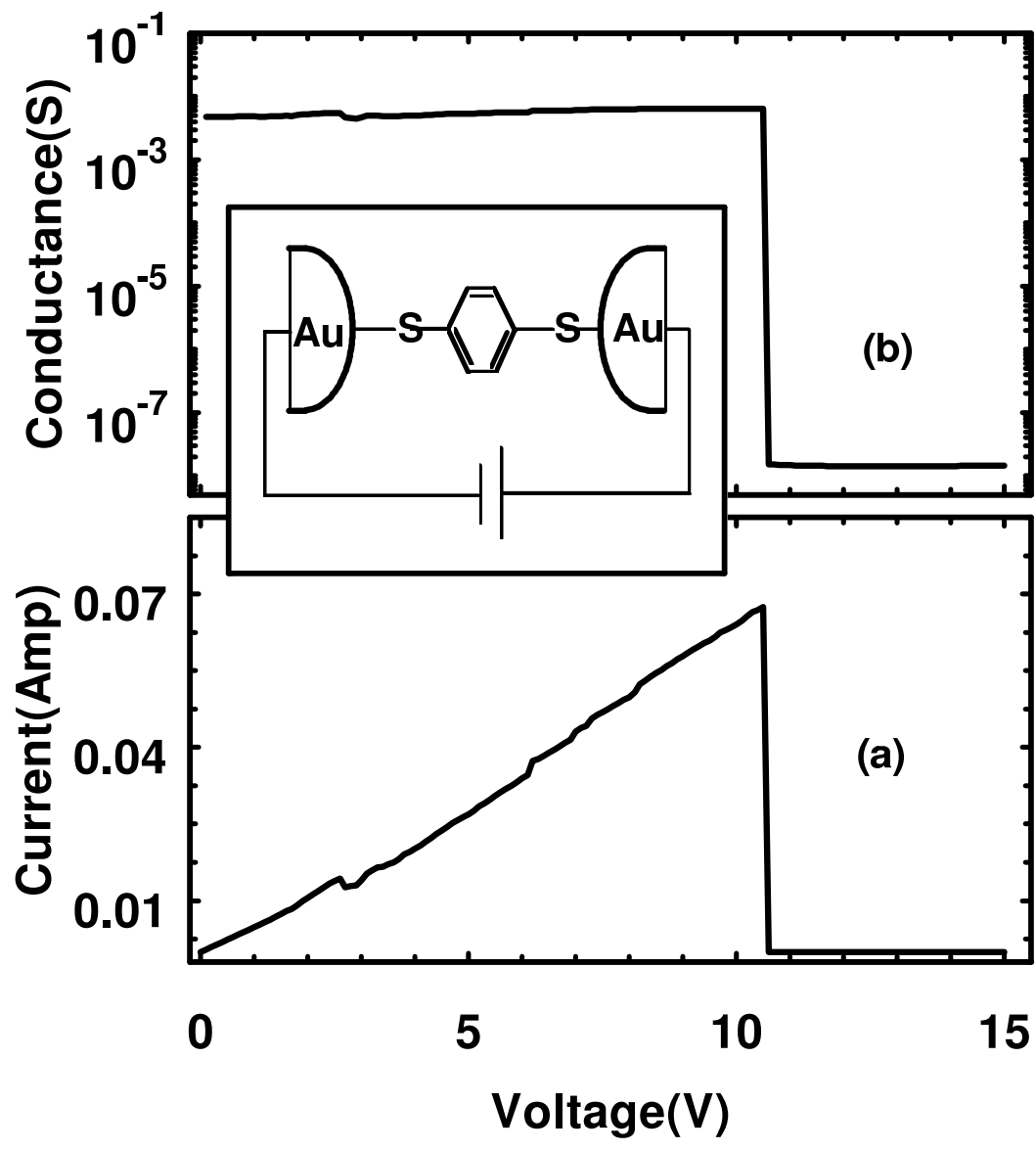
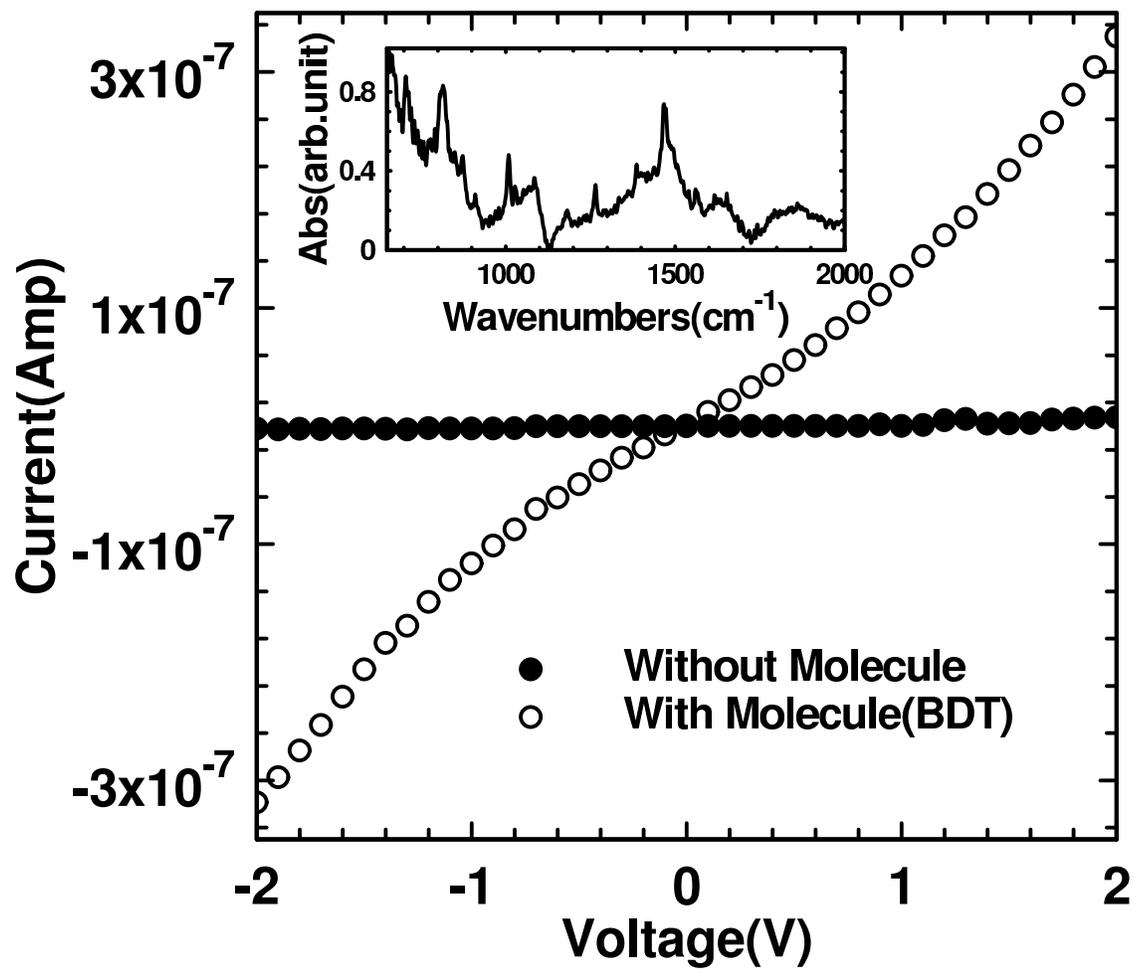
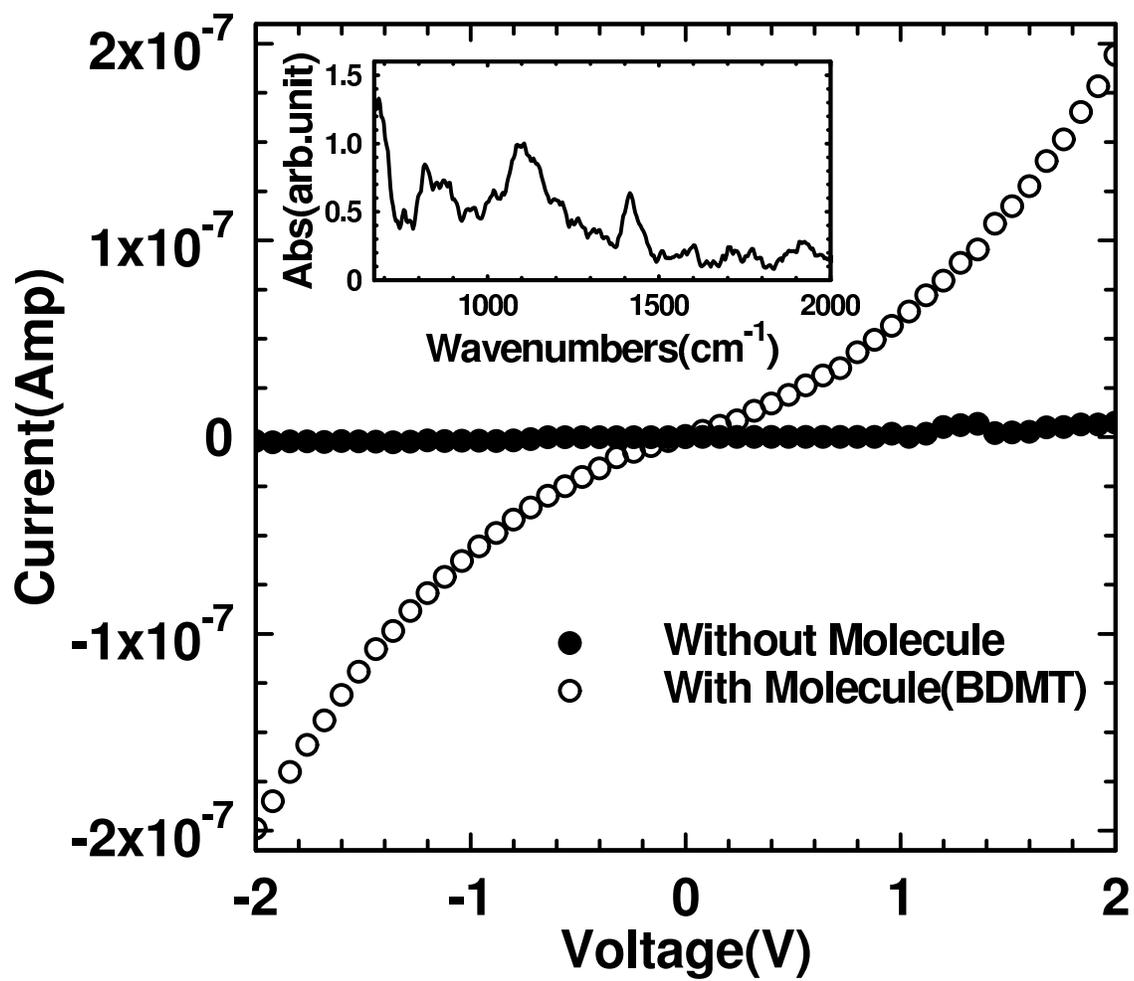
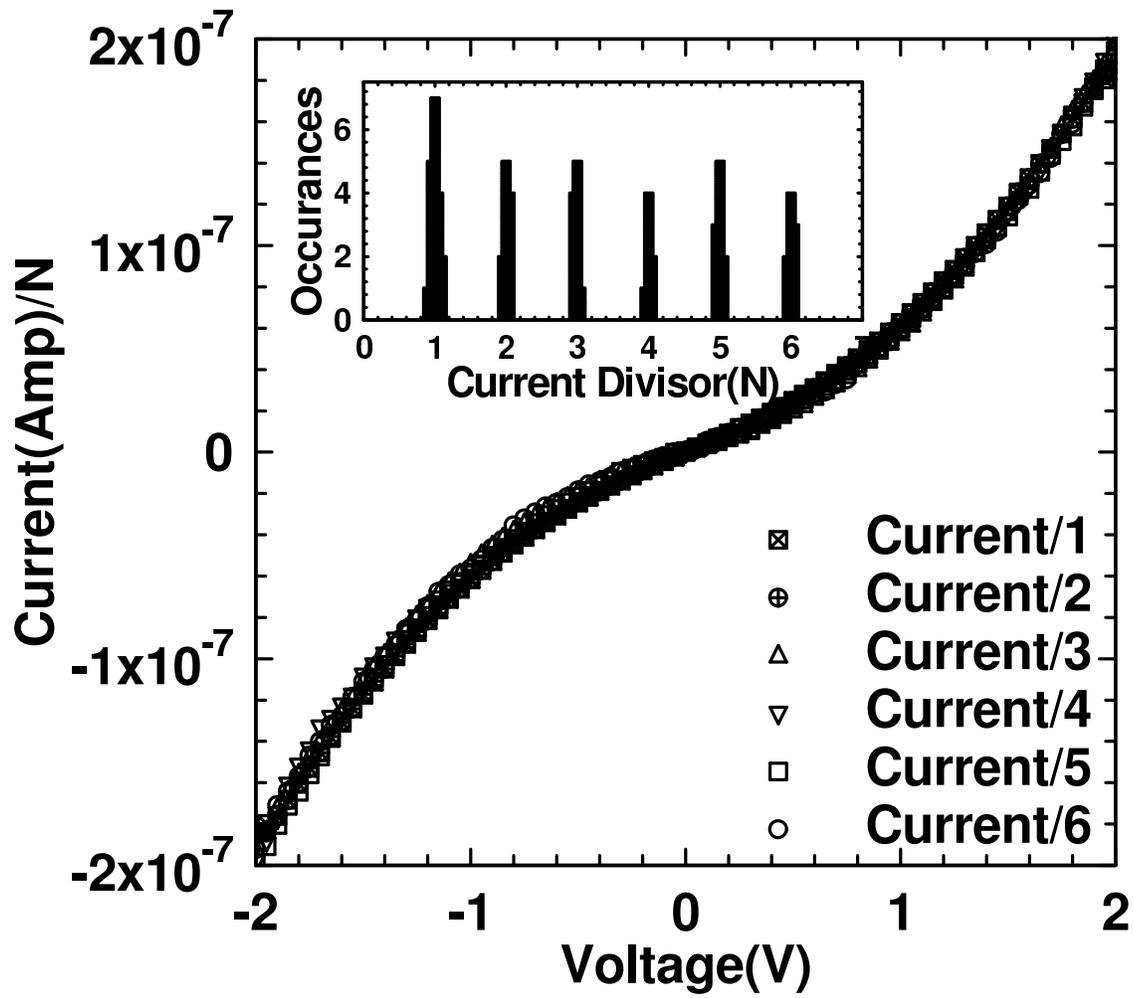
\end{document}